\documentclass[12pt]{article}
\begin{document}
\pagenumbering{arabic}
\title{Electrodynamics in accelerated frames revisited}

\author{J. W. Maluf$\,^{(a)}$  \\
Instituto de F\'{\i}sica, \\
Universidade de Bras\'{\i}lia\\
C. P. 04385 \\
70.919-970 Bras\'{\i}lia DF, Brazil\\
and \\
S. C. Ulhoa$\,^{(b)}$\\
Instituto de Ci\^encia e Tecnologia,\\
Universidade Federal dos Vales do Jequitinhonha e Mucuri,\\
UFVJM, Campus JK, Alto da Jacuba,\\
39.100-000 Diamantina, MG, Brazil}

\date{}
\maketitle
\begin{abstract}
Maxwell's equations are formulated in arbitrary moving frames by means of
tetrad fields, which are interpreted as reference frames adapted to observers
in space-time. We assume the existence of a general distribution of charges 
and currents in an inertial frame. Tetrad fields are used to
project the electromagnetic fields and sources on accelerated frames. 
The purpose is to study several configurations of fields and observers that in
the literature are understood as paradoxes. For instance, are the two 
situations, (i) an accelerated charge in an inertial frame, and (ii) a charge 
at rest in an inertial frame described from the perspective of an accelerated
frame, physically equivalent? Is the 
electromagnetic radiation the same in both frames? Normally in the 
analysis of these paradoxes the electromagnetic fields are transformed to 
(uniformly) accelerated frames by means of a coordinate transformation of the 
Faraday tensor. In the present approach coordinate and frame transformations 
are disentangled, and the electromagnetic field in the accelerated frame is
obtained through a frame (local Lorentz) transformation. Consequently the 
fields in the inertial and accelerated frames are described in the {\it same}
coordinate system. This feature allows the investigation of paradoxes such as
the one mentioned above.
\end{abstract}
\par
\bigskip
\noindent (a) wadih@unb.br, jwmaluf@gmail.com\par
\noindent (b) sc.ulhoa@gmail.com\par
\bigskip
\section{Introduction}
The electromagnetic theory defined by Maxwell's equations is a remarkable
theory developed more than a century ago. From the classical point of view,
the limits of the theory seem to be related to phenomena that involve 
electromagnetic radiation. The electromagnetic radiation emitted by a classical
electron in circular orbit is at the roots of the quantum theory. And the 
radiation of a linearly accelerated charged particle is a beautiful result of 
the theory that still nowadays is object of discussion. As viewed from a single
inertial frame, the electromagnetic radiation of an accelerated charged
particle is a well established result of the theory, except for the fact that 
so far it has not been verified experimentally. However, our intuition of this
phenomenon becomes less clear when we consider such radiation field from the 
point of view of an accelerated frame. Does an accelerated observer measure
electromagnetic radiation due to an equally accelerated charged particle? The
purpose of this paper is to try to answer this question, as well as to address
the two situations described in the Abstract, namely, (i) an accelerated 
charge in an inertial frame, and (ii) a charge at rest with respect to an 
accelerated frame. Is the electromagnetic radiation the same in both frames?

The electromagnetic field is described by the Faraday tensor $F^{\mu\nu}$. In
the present analysis we will consider that $\lbrace F^{\mu\nu}\rbrace$ are just
tensor components in the flat Minkowski space-time described by {\it arbitrary}
coordinates $x^\mu$. The projection of $F^{\mu\nu}$ on inertial or noninertial 
frames yield the electric and magnetic fields $E_x$, $E_y$, $E_z$, $B_x$, $B_y$
and $B_z$. The projection is carried out with the help of tetrad fields 
$e^a\,_\mu$. For instance, $E_x=-cF^{(0)(1)}$, where $c$ is the speed of light 
and $F^{(0)(1)}=e^{(0)}\,_\mu e^{(1)}\,_\nu F^{\mu\nu}$. 

Tetrad fields are considered as reference frames adapted to observers that 
follow trajectories described by functions $x^\mu(s)$ in space-time. These 
fields project vectors and tensors in space-time on the local frame of 
observers. The local projection of the vector $A^\mu(x)$ in space-time, for 
instance, is defined by $A^a(x)=e^a\,_\mu(x) A^\mu(x)$, and the projection
of the Faraday tensor is $F^{ab}(x)=e^a\,_\mu(x)e^a\,_\nu(x) F^{\mu\nu}(x)$.
Note that the right hand side and left hand side of these expressions are
evaluated at the same space-time event $x^\mu$. Therefore the projection is 
carried out in the same coordinate system. The measurable quantities are those
that are projected on the frame. Thus the laboratory quantities are $F^{ab}$.

In this paper we will write down equations for $F^{ab}$ that are completely 
equivalent to the well known Maxwell's equations. These equations hold in any
frame, inertial or noninertial frames. This formalism ensures that
the procedure for projecting electromagnetic fields on noninertial frames is
mathematically and physically consistent. Consequently we may investigate the
paradoxes mentioned above. The comparison of the electromagnetic fields in
inertial and noninertial frames is possible because these fields are defined in
the same coordinate system. We will conclude that the radiation of an 
accelerated charged particle in an inertial frame is different from the 
radiation of the charged particle at rest, as viewed from an equally 
accelerated frame. Consequently, the accelerated motion in space-time is not 
relative, and the radiation of an accelerated charged particle is an absolute
feature of the theory.

Electromagnetic radiation in accelerated systems has been addressed by Anderson
and Ryon \cite{Anderson}. They analyzed the three possible cases: I. observer
inertial, medium accelerated; II. observer accelerated, medium inertial; III.
observer and medium co-accelerated. The subject has also been investigated by
other authors \cite{Kovetz,Boulware,Eriksen,Saa}. A common feature to all these
approaches is that the accelerated frame is determined by means of a coordinate
transformation of the Faraday tensor. Therefore in these investigations
coordinate transformations and Lorentz transformations stand on equal footing.
This is not the point of view that we adopt in this paper. Coordinate
and Lorentz transformations are mathematically different transformations, and
we bring this difference to the physical realization of the theory.\par
\bigskip

Notation: space-time indices $\mu, \nu, ...$ and Lorentz (SO(3,1)) indices 
$a, b, ...$ run from 0 to 3. Time and space indices are indicated according to 
$\mu=0,i,\;\;a=(0),(i)$. The space-time is flat, and therefore the metric 
tensor is $g_{\mu\nu}= (-1,+1,+1,+1)$ in cartesian coordinates. The flat,
tangent space Minkowski space-time metric tensor raises and lowers tetrad 
indices and is fixed by 
$\eta_{ab}= e_{a\mu} e_{b\nu}g^{\mu\nu}=(-1,+1,+1,+1)$. The frame components
are given by the inverse tetrads $e_a\,^\mu$, although we may as well refer to
$\lbrace e^a\,_\mu\rbrace$ as the frame. The determinant of the tetrad 
field is represented by $e=\det(e^a\,_\mu)$.

\section{Reference frames in space-time}
Tetrad fields constitute a set of four orthonormal vectors in space-time,
$\lbrace e^{(0)}\,_\mu, e^{(1)}\,_\mu, e^{(2)}\,_\mu, e^{(3)}\,_\mu\rbrace$, 
that establish the local reference frame of an observer that moves along a 
trajectory $C$, represented by functions $x^\mu(s)$ 
\cite{Hehl1,Maluf1,Maluf2} ($s$ is the proper time of the observer). The tetrad
field yields the space-time metric tensor $g_{\mu\nu}$ by means of the relation
$e^a\,_\mu e^b\,_\nu \eta_{ab}=g_{\mu\nu}$, and $e^{(0)}\,_\mu$ and 
$e^{(i)}\,_\mu$ are timelike and spacelike vectors, respectively. We identify 
the timelike component of the frame with the observer's velocity 
$u^\mu=dx^\mu/ds$ along the trajectory: $e_{(0)}\,^\mu = u^\mu$.

The acceleration $a^\mu$ of the observer is given by the absolute derivative
of $u^\mu$ along $C$, 

\begin{equation}
a^\mu= {{Du^\mu}\over{ds}} ={{De_{(0)}\,^\mu}\over {ds}} =
u^\alpha \nabla_\alpha e_{(0)}\,^\mu\,, 
\label{1}
\end{equation}
where the covariant derivative is constructed out of the Christoffel symbols.
Thus the derivative of $e_{(0)}\,^\mu$ yields the acceleration 
along the worldline of an observer adapted to the frame. Therefore a set of 
tetrad fields for which $e_{(0)}\,^\mu$ describes a congruence of timelike 
curves is adapted to a class of observers characterized by the velocity field
$u^\mu=e_{(0)}\,^\mu$ and by the acceleration $a^\mu$. If 
$e^a\,_\mu = \delta^a_\mu$ everywhere in space-time, then $e^a\,_\mu$ is 
adapted to inertial observers, and $a^\mu=0$.

The acceleration of the whole frame is determined by the absolute derivative of
$e_a\,^\mu$ along $x^\mu(s)$. Thus, assuming that the observer carries an 
orthonormal tetrad frame $e_a\,^\mu$, the acceleration of the latter along the
path is given by \cite{Mashh1,Mashh2}

\begin{equation}
{{D e_a\,^\mu} \over {ds}}=\phi_a\,^b\,e_b\,^\mu\,,
\label{2}
\end{equation}
where $\phi_{ab}$ is the antisymmetric acceleration tensor. According to Refs.
\cite{Mashh1,Mashh2}, in analogy with the Faraday tensor we may identify
$\phi_{ab} \rightarrow ({\bf a}, {\bf \Omega})$, where ${\bf a}$ is the 
translational acceleration ($\phi_{(0)(i)}=a_{(i)}$) and ${\bf \Omega}$ is the
angular velocity of the local spatial frame with  respect to a nonrotating 
(Fermi-Walker transported \cite{Hehl1,Maluf2}) frame. It follows from Eq. (2)
that

\begin{equation}
\phi_a\,^b= e^b\,_\mu {{D e_a\,^\mu} \over {ds}}=
e^b\,_\mu \,u^\lambda\nabla_\lambda e_a\,^\mu\,.
\label{3}
\end{equation}
The accelerations $a^\mu$ and $\phi_{(0)(i)}$ are related via
$e^{(i)}\,_\mu a^\mu=e^{(i)}\,_\mu u^\alpha \nabla_\alpha
e_{(0)}\,^\mu=\phi_{(0)}\,^{(i)}$.

For a given frame determined by the set of tetrad fields $e^a\,_\mu$, the 
object of anholonomity $T^\lambda\,_{\mu\nu}$ is given by 
$T^\lambda\,_{\mu\nu}=e_a\,^\lambda T^a\,_{\mu\nu}$, where

\begin{equation}
T^a\,_{\mu\nu}=\partial_\mu e^a\,_\nu -\partial_\nu e^a\,_\mu\,.
\label{4}
\end{equation}
Note that $T^\lambda\,_{\mu\nu}$ is also the torsion tensor of the 
Weitzenb\"ock space-time. It is possible to show that in terms of 
$T^a\,_{\mu\nu}$ the acceleration tensor may be written as \cite{Maluf1,Maluf2}

\begin{equation}
\phi_{ab}={1\over 2} \lbrack T_{(0)ab}+T_{a(0)b}-T_{b(0)a}
\rbrack\,,
\label{5}
\end{equation}
where $T_{abc}=e_b\,^\mu e_c\,^\nu T_{a\mu\nu}$.

The expression for $\phi_{ab}$ is not covariant under local Lorentz (SO(3,1) or
frame) transformations, but is invariant under coordinate transformations.
The noncovariance under local Lorentz transformations allows us to take the
values of $\phi_{ab}$ to characterize the frame. The acceleration tensor
$\phi_{ab}$ represent the inertial accelerations on the frame along $x^\mu(s)$
\cite{Maluf1,Maluf2}. As an example, let us consider the tetrad fields adapted 
to observers at rest in Minkowski space-time. It is given by 
$e^a\,_\mu(ct,x,y,z)=\delta^a _\mu$. We then consider a time-dependent boost in
the $x$ direction, say, after which the tetrad field reads

\begin{equation}
e^a\,_\mu(ct,x,y,z)=\pmatrix{\gamma&-\beta\gamma&0&0\cr
-\beta\gamma&\gamma&0&0\cr
0&0&1&0\cr
0&0&0&1\cr}\,,
\label{6}
\end{equation}
where $\gamma=(1-\beta^2)^{-1/2}$, $\beta=v/c$ and $v=v(t)$. The frame above is
adapted to observers whose four-velocity is 
$u^\mu=e_{(0)}\,^\mu(ct,x,y,z)=(\gamma, \beta\gamma,0,0)$. After simple 
calculations we obtain \cite{Maluf1}

\begin{eqnarray}
\phi_{(0)(1)}&=&{d\over {dx^0}}\lbrack \beta \gamma\rbrack = 
{d \over {dt}}\biggl[
{ {v/c^2} \over {\sqrt{1-v^2/c^2} }} \biggr]\,, \\ \nonumber
\phi_{(0)(2)}&=&0 \,, \\ \nonumber
\phi_{(0)(3)}&=&0 \,,
\label{7}
\end{eqnarray}
and $\phi_{(i)(j)}=0$. The usual hyperbolic motion (uniform acceleration) is
characterized by $\phi_{(0)(1)}=a=$ constant.

For a static object whose four-velocity is given by
$V^\mu=(c,0,0,0)$ we may compute its frame components $V^a=e^a\,_\mu V^\mu$
with the help of eq. (6). We find $V^a=(\gamma c,-\beta\gamma c,0,0)$. Thus in
the classical limit $(v/c << 1)$ the velocity of the object with respect to the
accelerated frame is $V^{(1)}=-v(t)$, as expected.

\section{Maxwell's equations in moving frames}

Electrodynamics is formulated in terms of vector and tensor quantities, the
vector potential $A^\mu$ and the Faraday tensor $F^{\mu\nu}$ which are related
by $F_{\mu\nu}=\partial_\mu A_\nu - \partial_\nu A_\mu$. The sources are 
denoted by the four-vector current $J^\mu$. Space-time indices are raised and 
lowered by means of the flat space-time metric tensor 
$g_{\mu\nu}=(-1,+1,+1,+1)$. On a particular frame the electromagnetic 
quantities are projected according to $A^a(x)=e^a\,_\mu(x) A^\mu(x)$ and 
$F^{ab}(x)=e^a\,_\mu(x)e^a\,_\nu(x) F^{\mu\nu}(x)$. 

An inertial frame is 
characterized by the vanishing of the acceleration tensor $\phi_{ab}$. For 
instance, $e^a\,_\mu(t,x,y,z)=\delta^a_\mu$ describes an inertial frame because
it satisfies $\phi_{ab}=0$. More 
generally, all tetrad fields that are function of space-time {\it independent}
parameters (boost and rotation parameters) determine inertial frames. 
Suppose that $A^a$ are componentes of the vector potential in an inertial 
frame, i.e., $A^a=(e^a\,_\mu)_{in}A^\mu=\delta^a_\mu A^\mu$. The 
components of $A^a$ in a noninertial frame are obtained by means of a local
Lorentz transformation, 

\begin{equation}
\tilde{A}^a(x) = \Lambda^a\,_b(x) A^b(x)\,,
\label{8}
\end{equation}
where $\Lambda^a\,_b(x)$ are space-time dependent matrices that satisfy

\begin{equation}
\Lambda^a\,_c(x) \Lambda^b\,_d(x)\eta_{ab}=\eta_{cd}\,.
\label{9}
\end{equation}
Likewise, we have $\tilde{A}_a(x)=\Lambda_a\,^b(x)A_b(x)$. An alternative and
completely equivalent way of obtaining the field components $\tilde{A}_a(x)$
consists in performing a frame transformation by means of a suitable 
noninertial frame $e^a\,_\mu$, namely, in projecting $A^\mu$ on the noninertial
frame,

\begin{equation}
\tilde{A}^a(x)=e^a\,_\mu(x) A^\mu(x)\,.
\label{10}
\end{equation}

The covariant derivative of $A_a$ may be defined as

\begin{eqnarray}
D_a A_b&=&e_a\,^\mu D_\mu A_b \nonumber \\
&=&e_a\,^\mu (\partial_\mu A_b-{\;}^0\omega_\mu\,^c\,_b A_c)\,, 
\label{11}
\end{eqnarray}
where 

\begin{eqnarray}
^0\omega_{\mu ab}&=&-{1\over 2}e^c\,_\mu(
\Omega_{abc}-\Omega_{bac}-\Omega_{cab})\,, \\ \nonumber
\Omega_{abc}&=&e_{a\nu}(e_b\,^\mu\partial_\mu e_c\,^\nu-
                        e_c\,^\mu\partial_\mu e_b\,^\nu)\,,
\label{12}
\end{eqnarray}
is the metric-compatible Levi-Civita connection. Note that we are considering
the flat space-time, and yet this connection may be nonvanishing. In 
particular, for noninertial frames it is nonvanishing. The Weitzenb\"ock 
torsion tensor $T^a\,_{\mu\nu}$ is also nonvanishing. However, the
curvature tensor constructed out of ${}^0\omega_{\mu ab}$ vanishes:
$R^a\,_{b\mu\nu}({\,}^0\omega_{\mu ab})=0$. Under a local Lorentz 
transformation we have

\begin{equation}
\widetilde{^0\omega}_\mu\,^a\,_b=
\Lambda^a\,_c({\,}^0\omega_\mu\,^c\,_d)\Lambda_b\,^d
+\Lambda^a\,_c\partial_\mu \Lambda_b\,^c\,.
\label{13}
\end{equation}
It follows from eqs. (8), (12) and (13) that under a local Lorentz 
transformation we have

\begin{equation}
\tilde{D}_a\tilde{A}_b=\Lambda_a\,^c(x)\Lambda_b\,^d(x)\,D_cA_d\,.
\label{14}
\end{equation}

The Faraday tensor in a noninertial frame is defined as

\begin{equation}
F_{ab}=D_aA_b-D_bA_a\,.
\label{15}
\end{equation}
In view of eq. (14) we find that the tensors $F_{ab}$ and $\tilde{F}_{ab}$ 
in two arbitrary frames are related by

\begin{equation}
\tilde{F}_{ab}=\Lambda_a\,^c(x)\Lambda_b\,^d(x)F_{cd} \,.
\label{16}
\end{equation}

The Faraday tensor defined by eq. (15) is related to the standard expression
defined in inertial frames. By substituting (11) in (15) we find

\begin{eqnarray}
F_{ab}&=&e_a\,^\mu(\partial_\mu A_b-{\,}^0\omega_\mu\,^m\,_b A_m)-
e_b\,^\mu(\partial_\mu A_a-{\,}^0\omega_\mu\,^m\,_a A_m) \\ \nonumber
{}&=& e_a\,^\mu (\partial_\mu A_b)-e_b\,^\mu (\partial_\mu A_a)+
({\,}^0\omega_{abm}-{\,}^0\omega_{bam})A^m\,.
\label{17}
\end{eqnarray}
We make use of the {\it identity}

\begin{equation}
{\,}^0\omega_{abm}-{\,}^0\omega_{bam}=T_{mab}\,,
\label{18}
\end{equation}
where $T_{mab}$ is given by eq. (4), and write

\begin{eqnarray}
F_{ab}&=&e_a\,^\mu e_b\,^\nu(\partial_\mu A_\nu - \partial_\nu A_\mu)+
T^m\,_{ab}A_m \\ \nonumber
&{}&+e_a\,^\mu(\partial_\mu e_b\,^\nu)A_\nu-
e_b\,^\mu(\partial_\mu e_a\,^\nu)A_\nu\,.
\label{19}
\end{eqnarray}
In view of the orthogonality of the tetrad fields we have

\begin{equation}
\partial_\mu e_b\,^\nu=-e_b\,^\lambda(\partial_\mu e^c\,_\lambda)e_c\,^\nu\,.
\label{20}
\end{equation}
With the help of (20) we find that the last two terms of eq. (19) may be
rewritten as 

\begin{equation}
e_a\,^\mu(\partial_\mu e_b\,^\nu)A_\nu-
e_b\,^\mu(\partial_\mu e_a\,^\nu)A_\nu = -T^m\,_{ab}A_m\,.
\label{21}
\end{equation}
Therefore the last three terms of (19) cancel out and finally we have

\begin{equation}
F_{ab}=e_a\,^\mu e_b\,^\nu(\partial_\mu A_\nu - \partial_\nu A_\mu)\,.
\label{22}
\end{equation}

Thus $F_{ab}$ is just the projection of the Faraday tensor $F_{\mu\nu}$
in the noninertial frame determined by $e_a\,^\mu$. The scheme characterized
by eqs. (10-16) and (22) is in agreement with 
the procedure developed by Mashhoon 
\cite{Mashhoon} in the investigation of 
electrodynamics of accelerated systems, except that we deal with local fields,
contrary to Mashhoon, who considers a nonlocal representation of 
electromagnetic fields. A physical theory that is constructed out of local 
fields predicts phenomena whose measurements are pointwise. As argued by
Mashhoon, the Bohr-Rosenfled principle implies that only averages of field
components over a finite space-time region are physically meaningful, and
therefore a nonlocal formulation of electrodynamics is necessary for an
improvement of the theory. The nonlocal formulation of electrodynamics is 
still being developed, and does not seem to be mandatory in the present 
analysis. Note, however, that an ideal accelerated observer (to be discussed
in section 4) is described by a one-dimensional timelike trajectory in
space-time. Therefore the present formalism may admit nonlocality in time
(but not in space). The possible nonlocality in time will pose no problem to 
the analyses in section 4, since we will be interested in total quantities 
such as the total radiated power and the total radiated energy.

The covariant derivative of $F_{ab}$ is defined by

\begin{eqnarray}
D_a F_{bc}&=&e_a\,^\mu D_\mu F_{bc} \\ \nonumber
&=& e_a\,^\mu(\partial_\mu F_{bc}-{\,}^0\omega_\mu\,^m\,_b F_{mc}-
{\,}^0\omega_\mu\,^m\,_c F_{bm})\,.
\label{23}
\end{eqnarray}
Making extensive use of relations (18) and (20) we find that the source free 
Maxwell's equations in an arbitrary noninertial frame are given by

\begin{equation}
D_aF_{bc}+D_bF_{ca}+D_cF_{ab}= e_a\,^\mu e_b\,^\nu e_c\,^\lambda
(\partial_\mu F_{\nu\lambda}+\partial_\nu F_{\lambda \mu}+
\partial_\lambda F_{\mu\nu})=0\,.
\label{24}
\end{equation}

Maxwell's equations with sources are obtained from an action integral
whose Lagrangian density is given by

\begin{equation}
L=-{1\over 4} e\,F^{ab}F_{ab}- \mu_0\,e\,A_b J^b \,,
\label{25}
\end{equation}
where $e=\det(e^a\,_\mu)$, $J^b=e^b\,_\mu J^\mu$ and $\mu_0$ is the magnetic
permeability constant. Although in flat space-time we have $e=1$, we keep $e$
in the expressions below because it allows a straightforward inclusion of 
the gravitational field. Note that in view of eq. (22) we have

\begin{equation}
F^{ab}F_{ab}=F^{\mu\nu}F_{\mu\nu}\,,
\label{26}
\end{equation}
and therefore $L$ is frame independent. The field equations derived from $L$
are

\begin{equation}
\partial_\mu(e\,F^{\mu b})+ e\,F^{\mu c}\,({}^0\omega_\mu\,^b\,_c) 
=\mu_0\, e\,J^b\,,
\label{27}
\end{equation}
or

\begin{equation}
e_b\,^\nu \lbrack
\partial_\mu(e\,F^{\mu b})+ e\,F^{\mu c}\,({}^0\omega_\mu\,^b\,_c) \rbrack
=\mu_0\, e\,J^\nu\,,
\label{28}
\end{equation}
where $F^{\mu c}=e_b\,^\mu F^{bc}$. In view of eq. (26) it is clear that the 
equations above are equivalent to the standard form of Maxwell's equations in
flat space-time.

Equations (24) and (27) are equations for the electromagnetic field components
$F_{ab}$ in flat space-time, in arbitrary noninertial frames. They correspond
to projections of the standard Maxwell's equations on an
arbitrary frame determined by $e_a\,^\mu$.

The definition of a Lagrangian density such as eq. (25) is not unique. One 
could instead define the Faraday tensor as

\begin{eqnarray}
{\cal F}_{ab}&=&\partial_a\, A_b-\partial_b\, A_a \\ \nonumber
&=&e_a\,^\mu \partial_\mu (e_b\,^\nu A_\nu)-
e_b\,^\mu \partial_\mu (e_a\,^\nu A_\nu)\,.
\label{29}
\end{eqnarray}
Out of the expression above one would consider the Lagrangian density $L^\prime$
defined by 

\begin{equation}
L^\prime=-{1\over 4} e\,{\cal F}^{ab}{\cal F}_{ab}- \mu_0\,e\,A_b J^b \,,
\label{30}
\end{equation}
The field equations derived from $L^\prime$ read

\begin{equation}
\partial_\mu(e\,F^{\mu b}) =\mu_0\, e\,J^b\,.
\label{31}
\end{equation}
With the help of expression (20) we may rewrite the field equations above with
only space-time indices. It reads

\begin{equation}
\partial_\mu F^{\mu\nu}+{1\over 2}F^{\mu\lambda}T^\nu\,_{\mu\lambda}=
\mu_0 J^\nu\,.
\label{32}
\end{equation}
This is precisely the equation presented in ref. \cite{Hehl2} (eq. (B.4.33)) in
the analysis of Maxwell's equations in an arbitrary noninertial frame. In view
of the discussion above it is clear that eq. (32) is not just the projection 
of the standard form of Maxwell's equations on an arbitrary noninertial frame.
Moreover, eq. (22) does not hold in this framework. The field equation (27) is
derived from a Lagrangian density constructed out of $F^{ab}F_{ab}$ given by 
(26), and therefore it is clear that if we make $J^\mu=0$ 
everywhere in space-time we necessarily arrive at $F_{ab}=0$. On the other
hand, considering eq. (32), it is not immediately clear that $J^\mu=0$ implies
$F_{ab}=0$ in arbitrary noninertial frames for which 
$T^\nu\,_{\mu\lambda}\ne 0$. Equation (32) could lead to nontrivial vacuum 
solutions, which would be a very interesting but unexpected and improbable
result of the theory.

As a straightforward consequence of eq. (28), we consider the formulation of
Gauss law in the frame determined by eq. (6), where $v=v(t)$. We assume the
existence of the current $J^\mu=(c\rho({\bf r},t),0,0,0)$, where $c$ is the 
speed of light. In an inertial frame we have 
$(J^a)_{in}=\delta^a_\mu J^\mu=(c\rho,0,0,0)$.
We will denote $F^{ab}$ the components of the Faraday tensor in the 
accelerated frame, and (to simplify the notation) $F^{\mu\nu}$ the components
in the inertial frame where the source $\rho({\bf r},t)$ is defined.

Gauss law is obtained by taking the $\nu=0$ component of eq. (28). In view of
the notation above we have 

\begin{equation}
F^{ab}=\pmatrix{0&-{\tilde E_x}/c&-{\tilde E_y}/c&
-{\tilde E_z}/c\cr
{\tilde E_x}/c&0&-{\tilde B_z}&{\tilde B_y}\cr
{\tilde E_y}/c&{\tilde B_z}&0&-{\tilde B_x}\cr
{\tilde E_z}/c&-{\tilde B_y}&-{\tilde B_x}&0\cr}\,
\label{33}
\end{equation}
The components of $F^{\mu\nu}$ will be denoted without the tilde. The only
nonzero component of the Levi-Civita connection ${\,}^0\omega_{\mu ab}$ is
given by

\begin{equation}
{\,}^0\omega_{0(0)(1)}=
-{1\over c} \gamma^2 {{d \beta}\over {d t}}\,. 
\label{34}
\end{equation}
Substitution of (33) and (34) into the $\nu=0$ component of (28) yields,
after a number of simplifications,

\begin{equation}
\partial_x {\tilde E_x}+\gamma(\partial_y {\tilde E_y}+\partial_z {\tilde E_z})
+\beta c\gamma(\partial_y {\tilde B_z}-\partial_z {\tilde B_y})=
{\rho \over \varepsilon_0}\,.
\label{35}
\end{equation}

The electric field in the inertial frame is related, by means of local Lorentz
transformations, to the fields in the accelerated frame according to

\begin{eqnarray}
E_x&=&{\tilde E_x}\\ \nonumber
E_y&=&\gamma {\tilde E_y} +\beta c \gamma {\tilde B_z} \\ \nonumber
E_z&=&\gamma {\tilde E_z}-\beta c \gamma {\tilde B_y}\,.
\label{36}
\end{eqnarray}
After substitution of these expressions in eq. (35) we obtain the usual 
form of Gauss law $\nabla \cdot {\bf E}=\rho/\varepsilon_0$, as expected. 
Recall that $J^\mu=(c\rho({\bf r},t),0,0,0)$, and consequently 
$J^a=(\gamma c \rho, -\beta c \gamma \rho,0,0)$, by means of eq. (6).

We may instead consider the charge density to be ``at rest" in the accelerated
frame. In this case $J^a=(c\rho,0,0,0)$, which is obtained from
$J^\mu=(\gamma c \rho,\beta c \gamma \rho,0,0)$, and therefore Gauss law in the
accelerated frame in which the charge density is ``at rest" (i.e., the charge 
density is accelerated with respect to the inertial frame at rest) reads

\begin{equation}
\partial_x {\tilde E_x}+\gamma(\partial_y {\tilde E_y}+\partial_z {\tilde E_z})
+\beta c\gamma(\partial_y {\tilde B_z}-\partial_z {\tilde B_y})=
{{ \gamma \rho} \over \varepsilon_0}\,,
\label{37}
\end{equation}
which is similar to eq. (35), except that the charge density $\rho$ is 
increased by a factor $\gamma$. Written in terms of the inertial frame
components, eq. (37) reads $\nabla \cdot {\bf E}=(\gamma\rho)/\varepsilon_0$.

\section{Electromagnetic radiation in accelerated frames}

An ideal observer in space-time is defined by a timelike trajectory 
$x^\mu(s)$, where $s$ is the proper time, and $u^\mu=dx^\mu/ds$ is the 
observer's velocity. Thus the (one-dimensional) four-velocity 
$e_{(0)}\,^\mu=u^\mu$ describes the observer, and $e_a\,^\mu$ describes the 
whole frame. We assume that such ideal observer is equipped with gyroscopes 
that determine the orientation of the frame and with instruments that perform
pointwise measurements. The representation of the observer by a single world
line allows to simplify the analysis, and is not a fundamental limitation.
We will be ultimately interested in total values of field quantities such as 
the total radiated power, and thus the present setting is suitable for 
addressing the qualitative differences that arise in the calculations carried
out in inertial and noninertial frames.

The electric and magnetic field components $({\bf E}, {\bf B})$ and 
$({\bf \tilde{E}}, {\bf \tilde{B}})$ in the inertial and accelerated frames,
respectively, are related through the expression 
$F^{ab}=e^a\,_\mu e^b\,_\nu F^{\mu\nu}$, where $e^a\,_\mu$ is given by eq. (6).
The relations read

\begin{eqnarray}
\tilde{E}_x&=& E_x \,,\\ \nonumber
\tilde{E}_y&=& \gamma E_y -\beta c \gamma B_z \,, \\ \nonumber
\tilde{E}_z&=& \gamma E_z +\beta c \gamma B_y \,, \\ \nonumber
\tilde{B}_x&=& B_x \,, \\ \nonumber
\tilde{B}_y&=& \gamma B_y + {1\over c} \beta \gamma  E_z \,, \\ \nonumber
\tilde{B}_z&=& \gamma B_z - {1\over c} \beta \gamma  E_y  \,.
\label{38}
\end{eqnarray}
These relations will be used in the consideration of two known configurations
of electromagnetic fields. 

\subsection{An accelerated point charge in an inertial frame}

The first configuration is the field of an accelerated charged particle. Let
$x(t)$ represent the trajectory of a particle of charge $q$ restricted to move 
along the $x$ direction in an inertial frame. We define

\begin{equation}
{\bf b}(t)={{{\bf v}(t)}\over c}=
{1\over c}{{d x(t)}\over {dt}}{\bf \hat{x}}\,,
\label{39}
\end{equation}
such that ${\bf \dot{b}}\ne 0$. The point of observation in space is 
denoted by ${\bf r}$. We also define the vector

$${\bf R}(t)={\bf r} - x(t) {\bf \hat{x}}\,,$$
and ${\bf \hat{R}}={\bf R}/R$.
The electric and magnetic fields at the space-time event 
$({\bf r},t)$ are given by (see, for instance, Ref. \cite{Brau})

\begin{equation}
{\bf E}({\bf r},t)=
{q \over {4\pi \varepsilon_0}}\biggl[
{{{\bf \hat{R}}-{\bf b}}\over
{\gamma^2 R^2(1-{\bf b}\cdot{\bf \hat{R}})^3}}+
{{{\bf \hat{R}}\times\lbrack ({\bf \hat{R}}-
{\bf b})\times {\bf \dot{b}}\rbrack}\over 
{cR (1-{\bf b}\cdot{\bf \hat{R}})^3}}\biggr]_{t^\prime}\,,
\label{40}
\end{equation}

\begin{equation}
{\bf B}({\bf r},t)=
{1\over c}\lbrack {\bf \hat{R}}\times {\bf E}\rbrack_{t^\prime}\,,
\label{41}
\end{equation}
where $t^\prime$ is the retarded time, obtained as the solution of the equation

$$t^\prime =
t-{1\over c}\left| {\bf r}-x(t^\prime){\bf \hat{x}}\right| \,.$$

The frame will be co-moving with the accelerated charged particle, i.e., the
frame and the charged particle will be equally accelerated, if we require
the vector ${\bf b}(t)$ in eqs. (40) and (41) and $\beta(t)$ in eq. (38) to
satisfy $\left| {\bf b}(t) \right|=\beta(t)$, so that the charged 
particle will be at rest in the accelerated frame. It is clear that the
magnetic field ${\bf \tilde{B}}$, calculated out of (38) and (41), does not 
vanish in the accelerated frame (it can be easily calculated), and both 
$({\bf \tilde{E}}, {\bf \tilde{B}})$ generate a nontrivial Poynting
vector $\bf \tilde{S}$. The total power radiated by the point charge is
nonvanishing in the co-moving frame. 

Note that the Poynting vector is related to the $T^{0i}$
components of the energy-momentum tensor $T^{\mu\nu}$ of the electromagnetic
field \cite{Landau}. For arbitrary components of $T^{\mu\nu}$,
a frame transformation defined by eq. (6) (or a local Lorentz transformation)
in general leads to nonvanishing $T^{(0)(i)}$ components. For instance, for
the $T^{(0)(1)}$ component we have

$$T^{(0)(1)}={{1+\beta^2}\over{1-\beta^2}}T^{01}-
\beta\,\gamma^2(T^{00}+T^{11})\,.$$
The right hand side of the expression above is clearly nonvanishing in the
limit $\beta<<1$.

In the analysis above we have assumed that the interval between the
point charge and a particular observer (both accelerated) is timelike, and that
they are not separated by a horizon. If the interval is spacelike, the observer
will not detect radiation. 

\subsection{A point charge at rest observed from the point of view of
an accelerated frame}

The second configuration of electromagnetic field consists in the field of a
point charge at rest, at the origin (say) of an inertial frame. It generates
only the Coulomb field ${\bf E}=(E_x, E_y, E_z)$. The electric
field ${\bf E}$ varies with the radial distance as $1/r^2$. Let 
$({\bf \tilde{E}},{\bf \tilde{B}})$ represent the fields obtained in the
accelerated frame by means of eq. (6) and of 
$F^{ab}=e^a\,_\mu e^b\,_\nu F^{\mu\nu}$. The Poynting vector in the 
accelerated frame is 

\begin{equation}
{\bf \tilde{S}}={1\over \mu_0}{\bf \tilde{E}}\times {\bf \tilde{B}}\,,
\label{42}
\end{equation}
whose components are given by

\begin{eqnarray}
\tilde{S}_x&=&-{1\over {\mu_0 c}}\beta \gamma^2(E_y^2+E_z^2) \\ \nonumber
\tilde{S}_y&=&-{1\over {\mu_0 c}}\beta \gamma\,E_xE_y \\ \nonumber  
\tilde{S}_z&=& {1\over {\mu_0 c}}\beta \gamma\,E_x E_z\,,
\label{43}
\end{eqnarray}
in view of (38). It is clear from the expressions above that the Poynting 
vector ${\bf \tilde{S}}$ varies with the radial distance as $1/r^4$, and 
therefore the total power due to ${\bf \tilde{S}}$, measured in the accelerated
frame, vanishes. Thus this situation is not physically equivalent to that in 
which the point charge is accelerated with respect to an inertial frame. The
two situations are not relative to each other.

The difference between the two physical situations discussed above becomes more
clear in the nonrelativistic limit where $v(t)$ is finite but $\beta <<1$. For 
the two physical situations the integral of the Poynting vector over a 
two-dimensional spherical surface of constant radius $r_0$ around the observer 
can be easily calculated and compared to each other. Note that the tetrad field
for an inertial observer $e^a\,_\mu(t,x,y,z)=\delta^a_\mu$ and for the 
accelerated observer given by eq. (6) are written in the same coordinate 
system, and therefore a (spacelike) spherical surface of constant radius may be
taken to be the same for both observers. In the evaluation of total quantities
we require $r_0\rightarrow \infty$.

\section{Conclusion}

In this paper we have investigated the formulation of electrodynamics in 
accelerated frames in flat space-time. 
The Faraday tensor and Maxwell's equations are 
considered as vector and tensor quantities in the space-time described by
arbitrary coordinates $\lbrace x^\mu\rbrace$, and are projected on the frame
of an accelerated observer by means of tetrad fields. We are then able to 
obtain a consistent formulation of Maxwell's equations in any
noninertial frame in flat space-time. The advantage of our approach is that
the Faraday tensor (as well as Maxwell's equations) in the inertial and
noninertial frames are written in the same coordinate system, a feature that 
allows the comparison of the fields in the two frames.

The introduction of the gravitational field is straightforward. It amounts to
replacing the flat space-time tetrad field by the one that yields the 
gravitational field according to $e^a\,_\mu e_{a\nu}=g_{\mu\nu}$, and that is 
adapted to an observer, as described in section 2. The tetrad field describes
both a noninertial frame and the gravitational field.

The conclusion of the two situations discussed in the section 4 is that 
the accelerated motion in space-time is intrinsically absolute, not 
relative. The accelerated motion of a point charge in an inertial frame is not
physically equivalent to a point charge at rest with respect to an accelerated 
frame. Moreover, the radiation field of an accelerated point charge is 
measurable even in a co-moving frame. The relative motion in space-time seems 
to be verified only in the realm of inertial frames in Special Relativity. This 
conclusion holds as long as the interpretation of the tetrad field as a 
geometrical quantity that projects vectors and tensors on frames is valid.

Equations (24) and (27) may be worked out to yield the equations for 
electromagnetic waves in arbitrary accelerated frames. This issue will be 
investigated elsewhere.

\par
\bigskip
\noindent {\bf Acknowledgement}\par
\noindent We are grateful to C. A.P. Galv\~ao for bringing to our attention
\cite{Anderson} and \cite{Kovetz}.

\end{document}